\documentstyle[amssymb]{amsart}
\newtheorem{theorem}{Theorem}[section]
\newtheorem{lemma}[theorem]{Lemma}
\newtheorem{corollary}[theorem]{Corollary}
\theoremstyle{definition}
\newtheorem{definition}[theorem]{Definition}
\newtheorem{definitions}[theorem]{Definitions}
\theoremstyle{remark}
\newtheorem{fact}{Fact}  
\newcommand{\pr}{\operatorname{pr}}

\newcommand{\ex}{\bold{E}}
\newcommand{\rvx}{\bold{X}}
\newcommand{\rvy}{\bold{Y}}
\newcommand{\rvc}{\bold{C}}

\begin{document}
\title[The Threshold of Chaos in Random Boolean Nets]
{On the Threshold of Chaos in Random Boolean Cellular Automata}
\author{James F. Lynch}
\address{Department of Mathematics and Computer Science \\
Clarkson University \\
Potsdam, N. Y. 13699-5815}
\email{jlynch@@sun.mcs.clarkson.edu}
\thanks{Research supported by NSF Grant CCR-9006303.}
\maketitle
\begin{abstract}
A random boolean cellular automaton is a network of boolean gates
where the inputs, the boolean function, and the initial state of each
gate are chosen randomly. In this article, each gate has two inputs.
Let $a$ (respectively $c$) be the probability the the gate
is assigned a constant function (respectively a non-canalyzing
function, i.e., {\sc equivalence} or {\sc exclusive or}). Previous work
has shown that when $a > c$, with probability asymptotic to 1,
the random automaton exhibits very stable
behavior: almost all of the gates stabilize, almost all of them are
weak, i.e., they
can be perturbed without affecting the state cycle that is entered,
and the state cycle is bounded in size. This article gives evidence
that the condition $a = c$ is a threshold of chaotic behavior:
with probability asymptotic to 1,
almost all of the gates are still stable and weak, but
the state cycle size is unbounded. In fact, the average
state cycle size is superpolynomial in the number of gates.
\end{abstract}
\section{Introduction}
A topic of current interest in the theory of complex systems is the
existence of sharp boundaries between highly ordered and chaotic
behavior. Evidence for this phenomenon has been provided by computer
simulations, where some parameter is varied. As the parameter passes
through a certain critical region, the behavior of the system rapidly
changes between the two extremes of stability and chaos \cite{lang}.
In this article, we examine one of the simplest, yet most intensively
studied, models of complex systems --- the random boolean cellular automaton.
We present analytic results proving that there is such a threshold for
these systems.

Boolean cellular automata were introduced by Kauffman in \cite{k.1}.
He was interested in determining the conditions when complex systems
exhibit stable behavior. Three ways of measuring the stability are:
\begin{enumerate}
\item The proportion of gates that stabilize, i.e. eventually stop
changing.
\item The proportion of weak gates, i.e., gates that can be perturbed without
affecting the state cycle that is entered.
\item The size of the state cycle that the system eventually enters.
\end{enumerate}
The second and third of these measures are finite discrete analogues of
criteria that are used to characterize chaos in dynamical systems.
A small proportion of weak gates is similar to sensitivity to initial
conditions, and a large state cycle is similar to nonperiodicity.

Computer simulations, beginning with those described in \cite{k.1}, have
suggested that certain classes of randomly constructed boolean cellular
automata possess all three forms of stability with high probability. The basic
random model is where each gate has two inputs, and the inputs, the boolean
functions assigned to the gates, and the initial state are all chosen with
uniform probability distributions.  In particular, for each gate, each of the
16 boolean functions of two arguments has probability 1/16 of being assigned to
the gate.

In spite of extensive experimental work on these automata, comparatively little
has actually been proven about them. The first article containing formal
proofs of stability in the basic model is by {\L}uczak and Cohen \cite{lc}.
They show that as $n \rightarrow \infty$, for almost all random boolean
cellular automata with $n$ gates, the number of stable gates and the number of
weak gates is asymptotic to $n$. They also give a nontrivial upper bound on the
state cycle size. In Lynch \cite{l.rbca1}, it was shown that by giving a slight
bias to the probability of certain of the boolean functions assigned to the
gates (on the order of $\log\log n/\log n$), for almost all random boolean
cellular automata with $n$ gates, the state cycle size can be bounded above by
$n^{\gamma}$, for some $\gamma$. However, the proof failed when the bias was
reduced to 0, i.e. for the basic random model. This suggested two lines of
research. First, a more extensive analysis of random boolean cellular automata
with nonuniform probabilities of the boolean functions might be possible. This
could be a step toward understanding more realistic models of complex systems.
Also, the breakdown of the proof at the uniform distribution hinted at a
threshold phenomenon.

Treating all 16 of the two argument boolean functions individually seems to be
a complex undertaking. A classification of the boolean functions due to
Kauffman \cite{k.2} has proven useful. He referred to certain boolean
functions as canalyzing.
We will define this precisely in the next
section, but for now it suffices to note that among the canalyzing
functions are the constant functions; i.e. the function that outputs 0
regardless of its inputs and its negation that always outputs 1. Further,
among the two-argument boolean functions, there are only two non-canalyzing
functions: the {\sc equivalence} function that outputs 1 if and only if
both of
its inputs have the same value, and its negation the {\sc exclusive or}.

Let $a$ (respectively $c$) be the probability that the boolean function
assigned to a gate is constant (respectively noncanalyzing). In Lynch
\cite{l.rbca2} it was shown that when $a > c$,
with probability asymptotic to 1, the random boolean cellular
automaton is very stable in all three senses: almost all of the gates are
stable and weak, and the state cycle size is bounded.

In this article, we investigate the case $a = c \neq 0$. This includes the
basic model as the special case $a = c = 1/8$.
We prove that the first two
kinds of stability still hold (although the bounds here are not as tight), but
the state cycle size is unbounded for almost all automata. In fact,
the average state cycle size is greater than any polynomial in $n$. Thus, the
automaton still appears to be stable when viewed locally, i.e. at the level
of a typical gate, but large state cycles are a global symptom of the beginning
of instability.  In a future article, we will describe the behavior when
$a < c$.  At present, it is known that the proportion of weak gates is less
than $n$ by a nontrivial factor.
\section{Definitions}
Let $n$ be a natural number. A {\em boolean cellular automaton} $B$
with $n$ gates is a triple
$\langle D,F,x\rangle$ where $D$ is
a directed graph with vertices $1,\dots,n$ (referred to as {\em
gates}), $F = (f_1,\dots,f_n)$ is
a sequence of boolean functions, and $x = (x_1,\dots,x_n) \in \{0,1\}^n$
(the set of 0-1 sequences of length $n$). In this article, each gate
will have indegree two, and each boolean function will have two
arguments. We say that gate $j$ is an {\em input\/} to gate $i$ if
$(j,i)$ is an edge of $D$.
$B$ is a finite state automaton with state set $\{0,1\}^n$
and initial state $x$. The pair $\langle D,F\rangle$ defines the
transition function of $B$
in the following way. For each $i=1,\dots,n$ let
$j_i < k_i$ be the inputs of $i$.  Given $y =
(y_1,\dots,y_n) \in \{0,1\}^n$, $B(y) =
(f_1(y_{j_1},y_{k_1}),\dots,f_n(y_{j_n},y_{k_n}))$.
That is, the state of $B$ at time 0 is $x$, and if its
state at time $t$ is $y \in \{0,1\}^n$, then its state at time $t+1$ is
$B(y)$.

Our first set of definitions pertains to the aspects of
stability that will be studied.
\begin{definitions}\label{defstab}
Let $B = \langle D,F,x \rangle$ be a boolean cellular automaton.
\begin{enumerate}
\item We put $B^t(x)$ for the state of $B$ at time $t$, and
$f_i^t(x)$ for the value of its $i$th component, or gate, at time $t$.
\item Since the number of states is finite, i.e. $2^n$, there exist times
$t_0 < t_1$ such that $B^{t_0}(x) = B^{t_1}(x)$. Let $t_1$ be the
first time at which this occurs. Then $B^{t+t_1-t_0}(x) = B^t$ for all
$t \geq t_0$. We refer to the set of states $\{B^t(x) : t \geq t_0 \}$
as the {\em state cycle\/} of $B$, to
distinguish it from a cycle of $D$ in the graph-theoretic sense.
\item Gate $i$ {\em stabilizes\/} in $t$ steps
  if for all $t^\prime \ge
  t$, $f_i^{t^\prime}(x) = f_i^t(x)$.
\item Gate $i$ is {\em weak\/} if, letting $\overline{x}^i$
  be identical to $x$ except that its $i$th component is $1 - x_i$,
$$
{\boldsymbol \exists} t_0 {\boldsymbol \exists} d {\boldsymbol \forall} t
(t \ge t_0 \Rightarrow B^t(x) =
  B^{t + d}(\overline{x}^i)).
$$
That is, changing the state of $i$ does not affect the state cycle that
is entered.
\end{enumerate}
\end{definitions}

The next definitions describe a property of boolean functions that
plays a key role in the characterization of the threshold between
order and chaos.
\begin{definitions}
Let $f(x_1,x_2)$ be a boolean function of two arguments.
\begin{enumerate}
\item We say that $f$
{\em depends\/} on argument $x_1$ if for some $v \in \{0,1\}$,
$f(0,v) \neq f(1,v)$. A symmetric definition applies when $f$ depends on
$x_2$. Similarly, if $\langle D,F,x \rangle$ is a boolean
cellular automaton, $f_i = f$, and the inputs of gate $i$ are $j_{i1}$
and
$j_{i2}$, then for $m = 1,2$, $i$ depends on $j_{im}$ if $f$ depends on $x_m$.
\item
The function $f$ is said to be {\em canalyzing\/} if
there is some $m = 1$ or $2$ and some values $u,v \in \{0,1\}$ such that
for all $x_1,x_2 \in \{0,1\}$, if $x_m = u$ then $f(x_1,x_2) = v$.
Argument $x_m$ of $f$
is said to be a {\em forcing argument\/} with {\em forcing value\/} $u$ and
{\em forced value\/} $v$. Likewise, if $\langle D,F,x \rangle$
is a boolean
cellular automaton and $f_i$ is a canalyzing function with forcing argument
$x_m$, forcing value $u$ and forced value $v$, then input $j_{im}$ is a {\em
forcing input\/} of gate $i$.  That is, if the value of $j_{im}$ is $u$ at time
$t$, then the value of $i$ is guaranteed to be $v$ at time $t+1$.
\end{enumerate}
\end{definitions}

All of these definitions generalize immediately to boolean
functions of arbitrarily many arguments. In the case of two argument
boolean functions, the only non-canalyzing functions are
{\sc equivalence} and {\sc exclusive or}. The two constant functions
$f(x,y) = 0$ and $f(x,y) = 1$
are trivially canalyzing, as are the four functions that depend on only
one argument:
\begin{align*}
f(x,y) & =  x, \\
f(x,y) & =  \neg x, \\
f(x,y) & =  y \text{, and}\\
f(x,y) & =  \neg y.
\end{align*}
The remaining eight boolean functions of two arguments are canalyzing,
and they are all similar in the sense that both arguments are forcing with
a single value, and there is one forced value. A typical example is the
{\sc or} function. Both arguments are forcing with 1,
and the forced value is 1.

The notion of forcing, defined next, is a combinatorial condition that
is useful in characterizing stability. It depends on $D$ and $F$, but
not on $x$.
\begin{definition} Again, $\langle D,F,x \rangle$
is a boolean cellular automaton.
Using induction on $t$, we define what it means for gate $i$ to be {\em forced
to a value\/} $v$ {\em in\/} $t$ {\em steps}.

If $f_i$ is the constant function $f(x_1,x_2)=v$, then $i$ is forced to $v$ in
$t$ steps for all $t \ge 0$.

If the inputs $j_{i1}$ and $j_{i2}$ of $i$ are forced to $u_1$ and $u_2$
respectively in $t$ steps, then $i$ is forced to $f_i(u_1,u_2)$
in $t+1$ steps.

If $f_i$ is a canalyzing function with forcing argument $x_m$,
forcing value $u$,
and forced value $v$, and $j_{im}$ is forced to $u$ in $t$ steps, then
$i$ is forced to $v$ in $t+1$ steps.
\end{definition}
By induction on $t$ it can be seen that if $i$ is forced in $t$ steps, then
it stabilizes for all initial states $x$ in $t$ steps.

The following combinatorial notions will be used in characterizing
forcing structures. We assume the reader is familiar with the basic
concepts of graph theory (see e.g. Harary \cite{h}). Unless otherwise
stated, {\em path\/} and {\em cycle\/} shall mean directed path and
cycle in the digraph $D$.
\begin{definitions}
\begin{enumerate}
\item For any gate $i$ in $D$ with inputs
$j_{i1}$ and $j_{i2}$, let
\begin{align*}
S_0^-(i) & =  \{i\} \text{ and} \\
S_{d+1}^-(i) & =  S_d^-(j_{i1}) \cup S_d^-(j_{i2}).
\end{align*}
\item Then
$$
N_d^-(i) = \bigcup_{c \leq d} S_c^-(i).
$$
That is, $N_d^-(i)$ is the set of all gates that are connected to $i$ by
a path of length at most $d$.
\item If $I$ is a set of gates, then $N_d^-(I) = \cup_{i \in
I}N_d^-(i)$.
\item In a similar way we define $S_d^+(i)$ and $N_d^+(i)$, the set of all
gates reachable from $i$ by a path of length at most $d$.
\end{enumerate}
\end{definitions}
Note that whether $i$ is forced in $d$ steps is completely determined by
the
restriction of $D$ and $F$ to $N_d^-(i)$.

We will examine the asymptotic behavior of {\em random\/} boolean
cellular automata. For each boolean function $f$ of two arguments,
we associate a probability $a_f \in [0,1]$, where $\sum_f a_f = 1$.
The random boolean cellular automaton with $n$ gates is the result of
three random processes. First, a random digraph where every gate has indegree
two is generated. Independently for each gate, its two inputs are
selected from the \begin{math}
\bigl( \begin{smallmatrix}
n \\ 2
\end{smallmatrix} \bigr)
\end{math}
equally likely possibilities. Next, each gate is independently assigned a
boolean function of two arguments, using the probability distribution
$\langle a_f : f\colon\{0,1\}^2 \rightarrow \{0,1\}\rangle$. Lastly,
the initial state $x$ is chosen using the
uniform distribution on $\{0,1\}^n$.
We will use $\tilde{B} = \langle
\tilde{D},\tilde{F},\tilde{x}\rangle$ to denote a random boolean cellular
automaton generated as
above. For any properties $\cal P$ and $\cal Q$ pertaining to boolean cellular
automata, we put $\pr({\cal P},n)$ for the probability that the random boolean
cellular automaton on $n$ gates has property $\cal P$ and $\pr({\cal P}|
{\cal Q},n)$ for the conditional probability that $\cal P$ holds, given that
$\cal Q$ holds. Usually, we will omit the $n$ in these expressions since it
will be understood. Some of the properties we will investigate depend
only on $D$ and $F$. In that case, the expression describing $\cal P$
will involve $\langle \tilde{D},\tilde{F} \rangle$ instead of
$\tilde{B}$, and $\pr$ can be regarded as the probability measure
on random $\langle \tilde{D},\tilde{F} \rangle$. Similar notation
will be used for properties that depend only on $D$.
Random variables will be denoted by boldface
capital letters, and $\ex(\rvx)$ will be the expectation of $\rvx$.

We classify the two argument boolean functions as follows:
\begin{enumerate}
\item ${\cal A}$ contains the two constant functions.
\item ${\cal B}_1$ contains the four canalyzing functions that depend on
one argument.
\item ${\cal B}_2$ contains the eight canalyzing functions that depend on
both arguments.
\item ${\cal C}$ contains the two non-canalyzing functions.
\end{enumerate}
Then the probabilities that a gate is assigned a function in each of the
categories are:
\begin{align*}
a & =  \sum_{f \in {\cal A}} a_f \\
b_1 & =  \sum_{f \in {\cal B}_1} a_f \\
b_2 & =  \sum_{f \in {\cal B}_2} a_f \\
c & =  \sum_{f \in {\cal C}} a_f
\end{align*}
Lastly, we put ${\cal B} = {\cal B}_1 \cup {\cal B}_2$ and
$b = b_1 + b_2$, the probability that a gate is assigned a nonconstant
canalyzing function. Throughout the rest of the article, we assume the
following symmetry conditions on our distributions:
\begin{align*}
a_{f(x,y)} & = a_{f(y,x)} \text{ for all } f \in {\cal B}_1 \\
a_{f(x,y)} & = a_{f(\neg x,\neg y)} \text{ for all } f \in {\cal B}_2
\\
a_{f(x,y)} & = a_{\neg f(x,y)} \text{ for all } f \in {\cal C}.
\end{align*}
Also, $\log$ shall always mean $\log_2$, and $\ln$ is the natural
logarithm.
\section{Local Stability}
A key idea, first stated in \cite{lc}, is that almost all of the gates have
sufficiently large neighborhoods that are trees. We will use the following
version of this fact.
\begin{lemma}\label{lemloctree}
For any positive $\alpha$ and unbounded increasing function $\omega(n)$,
\begin{multline*}
\lim_{n \rightarrow \infty}\pr(\tilde{D} \text{ has at most }
\omega(n)(\log n)^3n^{2\alpha} \\
  \text{gates $i$ such that $N_{\alpha\log n}^-(i)$  is not a tree}) = 1.
\end{multline*}
The same is true for $N_{\alpha\log n}^+$.
\end{lemma}
\begin{pf}
For each gate $i$, let $\rvx_i$ be the indicator random variable that is 1
if and only if $N_{\alpha\log n}^-(i)$ is not a tree,
and let $\rvx = \sum_{i=1}^n \rvx_i$. If $\rvx_i = 1$, then
there exists a path P of length $p \le \alpha\log n$
beginning at some gate $k$ and ending at $i$
and another path $Q$ of length $q$, $1 \le q \le \alpha\log n$,
beginning at $k$,
disjoint from $P$ except at $k$ and its other endpoint,
which must be in
$P$. There are no more than $n^p$
ways of choosing $P$ and no more than $n^{q-1}\times p$ ways of choosing
$Q$. The probability of any such choice is bounded above by
$(2/n)^{p+q}$. Therefore
\begin{align*}
\ex(\rvx_i) & \le \sum_{p = 0}^{\alpha\log n} \sum_{q = 1}^{\alpha\log n}
2^{p+q}pn^{-1} \\
 & \le (\alpha\log n)^3 n^{2\alpha-1}.
\end{align*}
Then $\ex(\rvx) \le (\alpha\log n)^3 n^{2\alpha}$, and
the Lemma follows by Markov's inequality. A similar argument applies to
$N_{\alpha\log n}^+$.
\end{pf}

Another result we will need, from \cite{l.rbca2}, is a recurrence relation for
the probability that a gate is forced, given that its in-neighborhood is
treelike.
\begin{lemma}\label{lemlocrec}
For $d \ge 0$ and $v \in \{0,1\}$ let
\begin{align*}
p_d(v)  & =  \pr(\text{gate $i$ is forced to $v$ in $d$ steps } |
N_d^-(i) \text{ is a tree}) \text{ and} \\
p_d & = p_d(0) + p_d(1).
\end{align*}
Then
$$
p_d(0) = p_d(1)
$$
and $p_d$ satisfies the following recurrence.
\stepcounter{theorem}
\begin{align}
p_0 & =  a \text{ and} \notag\\
p_{d+1} & =  a + b p_d + c p_d^2.\label{eqrecp}
\end{align}
\end{lemma}

The fixed points of the recursion (\ref{eqrecp}) are $a/c$ and 1.
Consequently, when $a \ge c$, $p_d$ converges to 1. We will prove this for
$a = c$, but Figure 1 gives a graphical explanation of this fact.
Part (a) illustrates a typical case when $a > c$. In this case, as proven
in \cite{l.rbca2}, the convergence is geometric. The convergence when
$a = c$, shown in Part (b), is not as rapid, but is still sufficiently
fast.
\begin{lemma}\label{lemlocstab}
Let $d$ be a natural number. Then
$$
p_d \ge 1 - \frac1{ad}.
$$
\end{lemma}
\begin{pf}
Let $q_d = 1 - p_d$. Then from (\ref{eqrecp}), the recurrence for $q_d$ is
\stepcounter{theorem}\begin{equation}\label{eqrecq}
q_{d+1} = q_d - aq_d^2
\end{equation}
Letting $r_d = 1/q_d$ and using induction on $d$,
we will finish the proof by showing that $r_d \ge ad$. When $d = 0$, this is
evident. By (\ref{eqrecq}),
$$
\frac1{r_{d+1}} = \frac{1 - a/r_d}{r_d}
$$
and so
\begin{align*}
r_{d+1} &= \frac{r_d}{1 - a/r_d} \\
 & \ge r_d + a,
\end{align*}
which establishes the induction step.
\end{pf}

Our two main results on local stability are essentially generalizations of
similar results in \cite{lc}. Theorem \ref{thmweak} also improves the lower
bound on the number of weak gates that was given in \cite{lc}.
\vfill
\noindent
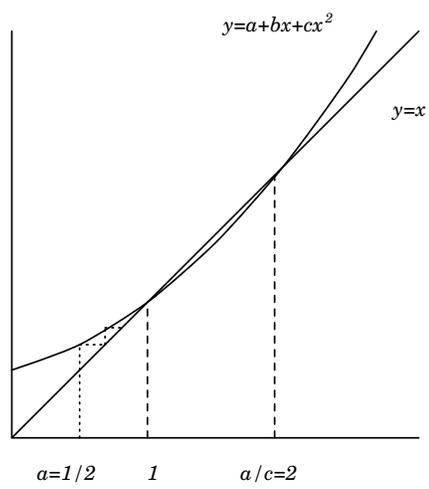
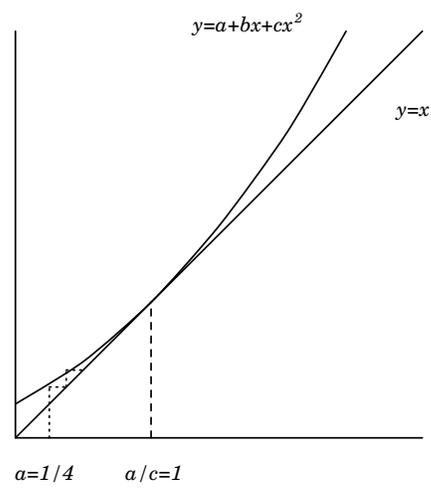
FIGURE 1. Examples of the convergence of $p_d$. The dotted line
$\cdot$$\cdot$$\cdot$$\cdot$
indicates the successive iterations of (\ref{eqrecp}) from $p_0 = a$
towards 1. \hfill \\ \hspace*{\fill} \\
(a) $a = 1/2$, $c = 1/4$. \hfill \\ \hspace*{\fill} \\
(b) $a = c = 1/4$.
\newpage
\begin{theorem}\label{thmstable}
Let $\alpha < 1/2$ and $\omega(n)$ be any unbounded increasing function.
Then
\begin{multline*}
\lim_{n \rightarrow \infty}
\pr(\langle\tilde{D},\tilde{F}\rangle
\text{ has at least } n(1 - \omega(n)/\log n) \\
  \text{gates that are forced in $\alpha\log n$ steps}) = 1.
\end{multline*}
\end{theorem}
\begin{pf}
Let $\rvy$ be the random variable that counts the number of gates $i$
in $\langle\tilde{D},\tilde{F}\rangle$
such that $N_{\alpha\log n}^-(i)$ is a tree
and $i$ is not forced in $\alpha\log n$ steps.
By Lemma \ref{lemlocstab},
$$
\ex(\rvy) \le \frac{n}{a\alpha\log n}.
$$
By Markov's inequality,
\begin{align*}
\pr\left(\rvy \ge \frac{n\omega(n)}{a\alpha\log n}\right) & \le
\frac1{\omega(n)} \\
 & \rightarrow 0.
\end{align*}
Therefore, together with Lemma \ref{lemloctree}, with probability asymptotic
to 1, there are at most
$$
\omega(n)\left[ \frac{n}{a\alpha\log n} + (\log n)^3n^{2\alpha}\right]
  = O\left(\frac{n\omega(n)}{\log n}\right)
$$
gates not forced in $\alpha\log n$ steps.
\end{pf}
Recalling that the notion of forcing is stronger than stability,
we have
\begin{corollary}
Let $\alpha < 1/2$ and $\omega(n)$ be any unbounded increasing function.
Then
\begin{multline*}
\lim_{n \rightarrow \infty}
\pr(\langle \tilde{D},\tilde{F} \rangle \text{ has at least }
  n(1 - \omega(n)/\log n) \\
  \text{gates that stabilize in $\alpha\log n$ steps}) = 1.
\end{multline*}
\end{corollary}

\begin{theorem}\label{thmweak}
Let $\omega(n)$ be any unbounded increasing function. Then
$$
\lim_{n \rightarrow \infty}\pr(\tilde{B} \text{ has at least
  $n(1 - \omega(n)/\log n)$
  weak gates}) = 1.
$$
\end{theorem}
\begin{pf}
We will use the following fact from \cite{lc}.
\begin{fact}\label{factlocn1}
For any gate $i$ and natural number $r$,
$$
\pr(|S_1^+(i)| = r) = \frac{2^r}{r!}e^{-2}
\left(1 + O\left(\frac{r}{n}\right)\right).
$$
\end{fact}
Thus, for $r > \log n$,
\begin{align*}
\pr(|S_1^+(i)| = r) & = O\left(\frac{(2e)^r}{r^r}\right) \\
 & = O(2^{-r\log r/2}) \\
 & = o(n^{-2})
\end{align*}
and the probability that there exists some gate with $|S_1^+(i)| > \log n$
is asymptotic to 0. For $r \le \log n$,
$$
\pr(|S_1^+(i)| = r) = \frac{2^r}{r!}e^{-2} + o(n^{-1/2}).
$$
By Lemma \ref{lemloctree}, this remains true even when the probability
is conditioned on $N_{\alpha\log n}^+(i)$ being a tree, $\alpha < 1/4$.

For any gate $i$ and natural number $d \le \alpha\log n$,
assuming $N_{\alpha\log n}^+(i)$ is a tree, let
$\phi_d$ be the probability that there is some gate
$j \in N_d^+(i)$ whose value is affected at step $d$, if the value of $i$ is
changed at step 0. That is, taking $\overline{x}^i$ as in Definitions
\ref{defstab} (4),
$f_j^d(\overline{x}^i) \neq f_j^d(x)$.
We will show by induction on $d = 1,\dots,\alpha\log n$
that $\phi_d \le 4/d$. Clearly $\phi_1 \le 1$.
Assuming $N_{\alpha\log n}^+(i)$ is a tree,
let $j \in S_1^+(i)$ and $\rho$ be the
probability that a change to $i$ affects $j$ in step 1. Since $N_{d+1}^+(i)$
is a tree, for any $k \in N_d^+(j)$, a change to $i$
affects $k$ in step $d+1$ if and only if a change to $i$ affects $j$
in step 1 and a change to $j$ affects $k$ in step $d$. Therefore, assuming
$|S_1^+(i)| \le \log n$,
\stepcounter{theorem}\begin{equation}\label{eqreqX}
\phi_{d+1} = 1 - \sum_{r=0}^{\lfloor \log n \rfloor}
  \pr(|S_1^+(i)| = r) \times
  [(1 - \rho) + \rho(1 - \phi_d)]^r.
\end{equation}

We show that $\rho = 1/2$. The three possibilities to consider are that
$f_j \in {\cal B}_1$, $f_j \in {\cal B}_2$, and $f_j \in {\cal C}$.
Let $k$ be the other input of $j$. Assuming $f_j \in {\cal B}_1$, two
out of the four functions in ${\cal B}_1$ result in $i$ affecting $j$ in
step 1. That is, if $i < k$ they are $f(x,y) = x$ and $f(x,y) = \neg x$,
and similarly for $k < i$. Altogether, the probability of the first case is
$b_1/2$ by the symmetry property $a_{f(x,y)} = a_{f(y,x)}$.
Now suppose $f_j \in {\cal B}_2$, and say $i < k$ and $x_k = 0$.
(The cases when $k < i$ or $x_k = 1$ are similar.)
Then $f_j(0,0) \neq f_j(1,0)$. But $f_j$ is canalyzing on both inputs,
so $f_j(0,1) = f_j(1,1)$. Four out of the eight functions in ${\cal B}_2$
satisfy these conditions, and the sum of their probabilities is
$b_2/2$ by the symmetry property $a_{f(x,y)} = a_{f(\neg x,\neg y)}$. The
probability of the third case is $c$, so altogether $\rho = b/2 + c = 1/2$.
Therefore by the Fact \ref{factlocn1} and Equation (\ref{eqreqX}),
\begin{align*}
\phi_{d+1} & = 1 - e^{-2} \sum_{r=0}^{\lfloor \log n
  \rfloor} \frac{(2 - \phi_d)^r}{r!} + o(n^{-1/2}\log n) \\
  & = 1 - e^{-\phi_d} + o(n^{-1/2}\log n) \\
  & \le \phi_d - \frac{\phi_d^2}2 + \frac{\phi_d^3}6 +
  o(n^{-1/2}\log n).
\end{align*}
If $\phi_d \le 1/(\log n)^2$, then $\phi_{d+1} \le 4/(\alpha\log n)
\le 4/(d+1)$. If $\phi_d > 1/(\log n)^2$, then
$\phi_{d+1} \le \phi_d - \phi_d^2/4$, and
using the same argument that was applied to Equation (\ref{eqrecq}),
$ \phi_{d+1} \le 4/(d+1)$.

Now let $\rvy$ be the random variable that counts the number of gates $i$
in $\tilde{B}$ such that $N_{\alpha\log n}^+(i)$ is a tree and $i$ is not
weak. Then by what we have just shown,
$$
\ex(\rvy) \le \frac{4n}{\alpha\log n}.
$$
The rest of the proof proceeds as in Theorem \ref{thmstable}.
\end{pf}
\section{Lower Bounds on Average State Cycle Size}
\subsection{Main Results}
Let the random variable $\rvc$ denote the size of the state cycle
of $\tilde{B}$.
\begin{theorem}\label{thmmain1}
For any constant $\gamma$ and sufficiently large $n$,
$$
\ex(\rvc) > n^{\gamma}.
$$
\end{theorem}
In the next theorem, $\ex(\rvc|\langle\tilde{D},\tilde{F}\rangle)$
is the expected state cycle size of a random $\langle\tilde{D},\tilde{F}
\rangle$ averaged over all $x \in \{0,1\}^n$.
\begin{theorem}\label{thmmain2}
There is a constant $\gamma > 0$ such that
$$
\lim_{n \rightarrow \infty} \pr(\ex(\rvc |
\langle\tilde{D},\tilde{F}\rangle) \ge n^{\gamma}) = 1.
$$
\end{theorem}

These theorems will follow from a key result (Lemma \ref{lemlbvortex}) on
the probability of existence of certain kinds of structures in
$\langle\tilde{D},\tilde{F}\rangle$.
We first define these structures and prove some basic facts about them. Let
$\alpha$ be a fixed real number such that $0 < \alpha < 1/2$. In the following
we will put $m$ for $\lceil\alpha \log n\rceil$.
\subsection{Vortices}
\begin{definition}\label{defvortex}
Let $B = \langle D,F,x\rangle$ be a boolean cellular automaton on $n$
gates. A {\em vortex\/} of circumference $d$ consists of two disjoint
subsets of gates $R = \{r_0,\dots,r_{d-1}\}$ and $S = \{s_0,\dots,s_{d-1}\}$
satisfying the following conditions for $0 \le i < d$.
\begin{enumerate}
\item $(r_i,r_{i+1 \pmod{d}}) \in D$.
\item $(s_i,r_i) \in D$.
\item $s_i$ is forced in $m$ steps.
\item The value that $s_i$ is forced to is not a forcing value for $f_{r_i}$.
\end{enumerate}
We refer to it as a vortex on $R,S$ or simply $R \cup S$ if we do not need to
distinguish $R$ and $S$.
\end{definition}
An example is given in Figure 2.

The essential characteristics of such a vortex are captured by the
directed labeled graph
$$
V = \langle R \cup S, D \upharpoonright (R \cup S),F \upharpoonright
  R,v_0,\dots,v_{d-1}\rangle
$$
where $v_i$ is the value that $s_i$ is forced to, for $i = 0,\dots,d-1$.
That is, $V$ is simply the restriction of $
\langle\tilde{D},\tilde{F}\rangle$ to $R \cup S$, with the
functions labeling the gates in $S$ replaced by their forced values.
The isomorphism class of $V$ is called a {\em vortex type}.

For any such vortex type $\tau$, and any $V \in \tau$ as above,
we put $\lambda(\tau)$ for the size of the automorphism group on $V$
and $\pi(\tau)$ for $\prod_{i = 0}^{d-1} a_{f_{r_i}}$. Clearly
$\lambda(\tau)$ and $\pi(\tau)$ do not depend on the choice of $V \in \tau$.
The significance of these two quantities is that $(2d)!/\lambda(\tau)$
is the number of distinct labelings of the gates in any $V \in \tau$,
and $\pi(\tau)$ is the conditional probability that two disjoint subsets
$R$ and $S$, each of size $d$, form a vortex of type $\tau$, given that
conditions (1)--(3) in Definition \ref{defvortex} hold.
The following two facts will be used later in the combinatorial analysis
of vortices. Let $T$ be the set of all vortex types of circumference $d$.
\newpage
\vspace*{\fill}
\noindent
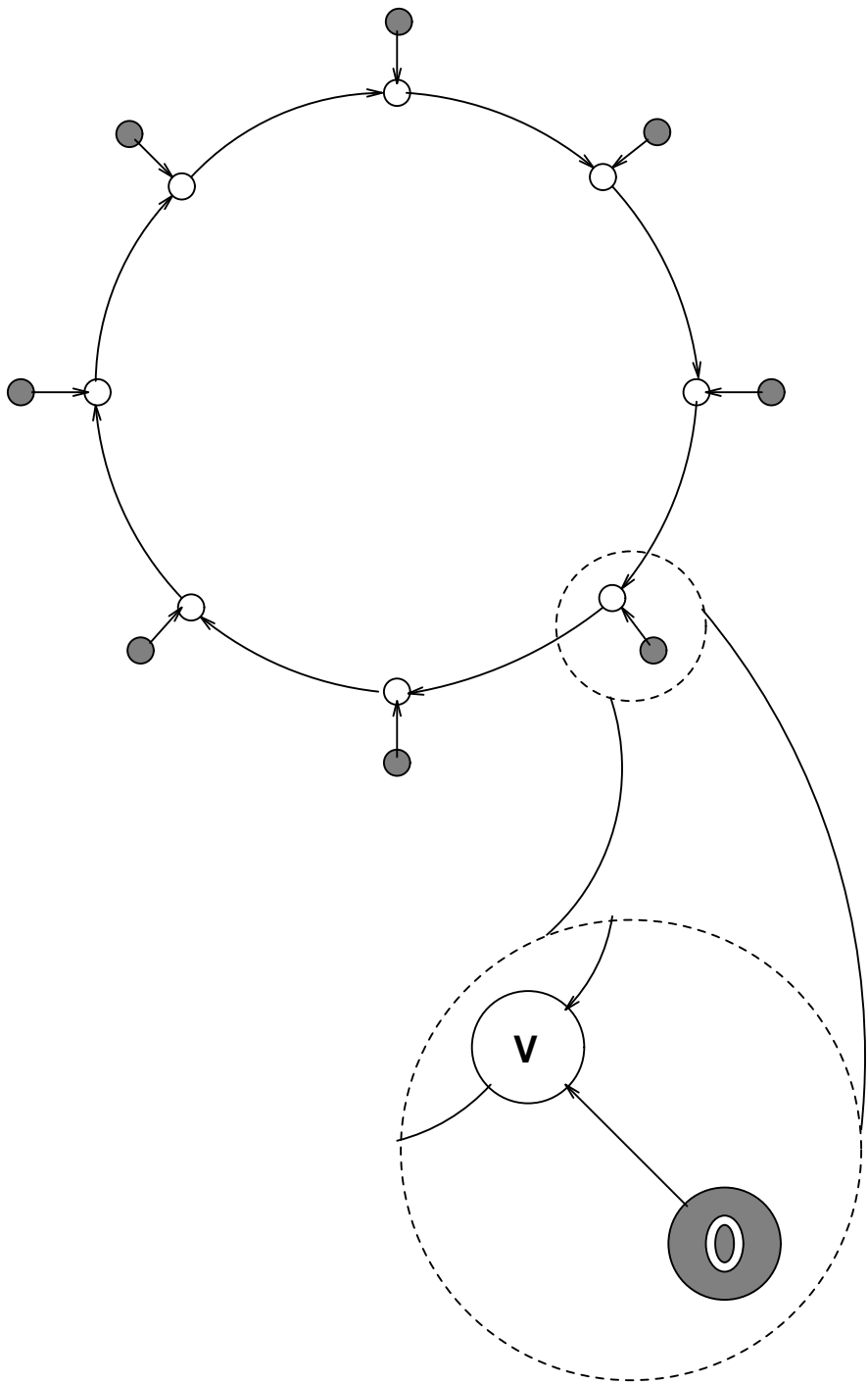
FIGURE 2. A schematic diagram of a vortex of circumference 8. Shaded
circles are members of $S$, and unshaded circles are in $R$. The
enlargement shows a typical $(s_i,r_i)$ pair. In this example, $s_i$
is forced to 0 while $f_{r_i} = \vee$ (the {\sc or} function).
\newpage
\begin{lemma}\label{lemlbrigid}
There exists $\rho \in (0,1)$ such that
$$
\sum\begin{Sb}\tau \in T \\
  \lambda(\tau) > 1\end{Sb} \pi(\tau) \le d\rho^{d/2}.
$$
\end{lemma}
\begin{pf}
The only nontrivial automorphisms of $V \in \tau$ are those that take each
$r_i$ to $r_{i+p \pmod{d}}$, where $1 \le p < d$. But this implies
\stepcounter{theorem}\begin{equation}\label{eqauto}
f_{r_i} = f_{r_{i+p \pmod{d}}} \text{ for } i = 0,\dots,d-1.
\end{equation}
We may assume $p$ is the minimal number satisfying (\ref{eqauto}),
and therefore $p | d$, so $p \le d/2$. Let
$\rho = \max\{a_f : f \notin {\cal A} \}$, $q = d/p$, and $T_p$ be the set
of vortex types satisfying (\ref{eqauto}). Then
\begin{align*}
\sum_{\tau \in T_p} \pi(\tau) & \le (\rho^p)^{q-1} \\
  & = \rho^{d-p} \\
  & \le \rho^{d/2}.
\end{align*}
The factor $d$ in the Lemma is a crude upper bound on the number of
divisors of $d$.
\end{pf}
\begin{lemma}\label{lemlbprob}
We have
$$
1 - d 2^{-d/2} \le \sum_{\tau \in T} \pi(\tau) \le 1.
$$
\end{lemma}
\begin{pf}
For any sequence $v = (v_0,\dots,v_{d-1}) \in \{0,1\}^d$, let $T_v$
be the set of all vortex types in $T$ such that the labeling of $S$ is
isomorphic to $v$. Let $U$ consist of all sequences $v \in \{0,1\}^d$
that do not have any nontrivial cyclic permutations, and let
$T^\prime = T - \cup_{v \in U}T_v$. Then, using the same methods as in
Lemma \ref{lemlbrigid}, $|U| \ge 2^d - d 2^{d/2}$. Since
$$
\sum_{\tau \in T}\pi(\tau) = \sum_{v \in U} \sum _{\tau \in T_v} \pi(\tau)
  + \sum_{\tau \in T^\prime} \pi(\tau),
$$
we will be done by showing that for all $v \in \{0,1\}^d$
\stepcounter{theorem}\begin{equation}\label{eqtau}
\sum_{\tau \in T_v}\pi(\tau) = 2^{-d}.
\end{equation}

For every $i = 0,\dots,d-1$, $v_i$ does not force $f_{r_i}$. Therefore one
of the following possibilities must hold.
\begin{enumerate}
\item $f_{r_i} \in {\cal B}_1$, and the input on which $r_i$ depends is
$r_{i-1 \pmod{d}}$.
\item $f_{r_i} \in {\cal B}_2$, and $v_i$ is not a forcing value for
$f_{r_i}$.
\item $f_{r_i} \in {\cal C}$.
\end{enumerate}
Case (1) has probability $b_1/2$ by the symmetry property
$a_{f(x,y)} = a_{f(y,x)}$, Case (2) has probability $b_2/2$ by the symmetry
property $a_{f(x,y)} = a_{f(\neg x,\neg y)}$, and Case (3) has probability
$c$. Therefore, given that $s_i$ is labeled with $v_i$, the probability
that one of the three cases above holds is $1/2$, and (\ref{eqtau}) follows.
\end{pf}

The existence of vortices of sufficiently large prime circumference will
be used to prove the lower bounds on average state cycle size. This is
the relevance of the next two basic facts. When we refer to the state of a
vortex, we simply mean the state of $B$ restricted to $R \cup S$.
\begin{lemma}\label{lemlbcycle}
A vortex enters its state cycle in at most $m$ steps. Its state cycle is
completely determined by the initial state of $R \cup N_m^-(S)$.
\end{lemma}
\begin{pf}
After $m$ steps, for each $i = 0,\dots,d-1$, $s_i$ is forced to some value
$v$. Since $v$ is not a forcing value for $f_{r_i}$, assuming $s_i <
r_{i-1 \pmod{d}}$ (the case when $s_i > r_{i-1 \pmod{d}}$ is symmetric),
$f_{r_i}(v,y) = y$ or $\neg y$. Let us use the notation $f_{r_i}(v,y) =
g_i(y)$ where $g_i(y) = y$ or $\neg y$, depending on which case holds.

In other words, after $m$ steps, the vortex is equivalent to a cycle of
1-input gates, none of which are constants. Let $u = (u_0,\dots,u_{d-1})$
be the state of these gates after $m$ steps. We need only show that $u$
reoccurs.

Suppose there is an even number of gates $r_i$ such that $g_i(y) = \neg y$.
Then after $m+d$ steps, the state of each $r_i$ will be $u_i$. If there is
an odd number, then the state of each $r_i$ after $m+2d$ steps will again
be $u_i$. In either case, the state cycle has been reentered in not more
than $2d$ steps.
\end{pf}
\begin{lemma}\label{lemlbprimevortex}
If the circumference of the vortex is prime, then the size of its state
cycle is 1, 2, $d$, or $2d$.
\end{lemma}
\begin{pf}
{}From the proof of Lemma \ref{lemlbcycle}, we know that the state repeats
every $2d$ steps, and thus the state cycle size is a factor of $2d$.
\end{pf}

To simplify calculations in the remainder of the proofs, we condition all
events on the following two properties. Let $\beta > \alpha$ be fixed.
\begin{enumerate}
\item There are no distinct vortices on $R,S$ and $R^\prime,S^\prime$
respectively of circumference less than or equal to $2\beta\log n$ such that
$$
(R \cup N_m^-(S)) \cap (R^\prime \cup N_m^-(S^\prime)) \neq \emptyset.
$$
\item For every vortex of circumference less than or equal to $2\beta\log n$
on any $R,S$, for all $s,s^\prime \in S$,
\begin{gather*}
N_m^-(s) \text{ is a tree,} \\
N_m^-(s) \cap N_m^-(s^\prime) = \emptyset \text{, and} \\
N_m^-(s) \cap R = \emptyset.
\end{gather*}
\end{enumerate}
A boolean cellular automaton satisfying these conditions is said to be
{\em simple}. By the next lemma, this will not affect the asymptotic
probabilities that will be computed.
\begin{lemma}\label{lemlbsimple}
We have
$$
\pr(\langle\tilde{D},\tilde{F}\rangle \text{ is simple}) = 1 - n^{-\Omega(1)}.
$$
\end{lemma}
\begin{pf}
One way that a boolean cellular automaton can fail Condition (1) above is if
there exist distinct vortices on $R,S$ and $R^\prime,S^\prime$ such that
$R \cap R^\prime \neq \emptyset$.
Then there are gates $r_i,r_j \in R$ (possibly the same) and a path of
gates in $R^\prime$ beginning at $r_i$ and ending at $r_j$, disjoint from
$R$ except at the endpoints. If the circumference of $R$ is $d$ and the length
of the path is $l$, then $p = d + l - 2$ is the number of gates in $R$
and the path. Letting $\kappa$ range over all choices of $(d,l,i,j,C)$
such that $d,l \leq 2\beta\log n$, $0 \le i,j < d$, and $C$ is a subset of
$\{1,\dots,n\}$ of size $p$, we put $\rvx_\kappa$ for the indicator random
variable that is 1 if and only if the gates in $C$ form a cycle $R$ and a
path as above. Then $\rvx = \sum_\kappa \rvx_\kappa$ is an upper bound on
the expected number of pairs of vortices such that $R \cap R^\prime \neq
\emptyset$.

Now
\begin{align*}
\ex(\rvx_\kappa) & \le p! \times \frac1{\binom{n}2}\times
  \left(\frac{n-1}{\binom{n}2}\right)^{p-1}
  \times \left(\frac12\right)^p \\
 & = \frac{p!}{(n-1)n^{p}}
\end{align*}
because $p!$ is an upper bound on the number labelings of $C$,
$1/\binom{n}2$ is the probability that Condition (1) of Definition
\ref{defvortex} holds for $r_j$,
$(n-1)/\binom{n}2$ is the probability that Condition (1) holds for all
other
gates in $C$, and $1/2$ is the probability that Condition (4) holds for
a gate, given
that Condition (3) holds. There are $O((\log n)^4)$ choices for $d$, $l$, $i$,
and $j$, and for each of these choices, there are $\binom{n}{p}$ choices for
$C$. Therefore
\begin{align*}
\ex(\rvx) & = O((\log n)^4 n^{-1}) \\
 & = n^{-\Omega(1)}.
\end{align*}

On the other hand, if $R \cap R^\prime = \emptyset$, but Condition (1) of
simplicity is still violated, then there exists a gate $g$ and two paths
$P$ and $P^\prime$ of lengths $p,p^\prime \le m+1$ beginning at $g$ and
disjoint everywhere else, one path ending in $R$ and the other in
$R^\prime$. There are at most $n$ ways of chosing $g$,
$(m+2)^2$ ways of choosing $p$ and $p^\prime$, and
$n^{p+p^\prime-2}\times(2\beta\log n)^2$ ways of choosing the remaining gates
in $P$ and $P^\prime$. The probability of such a choice is bounded above by
$(2/n)^{p+p^\prime}$. Therefore, by Markov's inequality, the probability that
$P$ and $P^\prime$ exist is bounded above by
\begin{align*}
n^{p+p^\prime-1}
\times (m+2)^2 \times (2\beta\log n)^2 \times
\left(\frac2{n}\right)^{p+p^\prime} & =
 O((\log n)^4 2^{2\alpha\log n} n^{-1}) \\
 & = n^{-\Omega(1)}.
\end{align*}

A similar proof enables us to show that Condition (2) holds with probability
$1 - n^{-\Omega(1)}$.
\end{pf}

One final condition on vortices that will be needed is that they should enter
a large (relative to their circumference) state cycle from many initial
states. This is formalized by the next definition.
\begin{definition}
A vortex of circumference $d$ is {\em strong} if for at least $1/2$ of the
initial states of $B$, the state cycle  of the vortex is greater than or
equal to $d$.
\end{definition}
\begin{lemma}\label{lemlbstrong}
If $\tilde{B}$ is simple, then for any
vortex $V$ of circumference $d \ge m+2$ where $d$ is prime, the
probability that $V$ is strong is greater than or equal to $1/2-o(1)$.
\end{lemma}
\begin{pf}
If at least $1/2$ of the initial states take $V$ to a state cycle of
size $\ge d$, then we are done. Otherwise, by Lemma \ref{lemlbprimevortex},
at least $1/2$ of the inputs take $V$ to a fixed point or a 2-cycle.

Since $d \ge m+2$, with probability $\ge 1 - o(1)$, $V$ has at least one
$i$ such that $f_{r_i} \in {\cal C}$. Without loss of generality, let us
assume $f_{r_0} \in {\cal C}$, and let $x$ be any input that takes $V$
to a fixed point or 2-cycle. Let $V^\prime$ be the vortex obtained from $V$
by changing $f_{r_0}$ to $\neg f_{r_0}$. Using the notation of Lemma
\ref{lemlbcycle},
this has the effect of changing
$g_0$ to $\neg g_0$.

Let $w_0$ and $w_1$ be the values of $r_{m+1}$ in $V$ at times $m$ and
$m+1$ respectively. Then, by Condition (2) in the definition of
simplicity, $w_0$ and $w_1$
are also the values of $r_{m+1}$ in $V^\prime$ at those times.
Since $V$ enters a fixed point or 2-cycle from $x$,
the sequence of values of $r_{m+1}$ beginning at time $m$ must be
$w_0,w_1,w_0,w_1,\dotsc$ (possibly $w_0 = w_1$). If $V^\prime$ also enters
a fixed point or 2-cycle, then the sequence of values of $r_{m+1}$ beginning
at $m$ is also $w_0,w_1,w_0,w_1,\dotsc$. In particular, assuming $m$ is even,
its state at time $2m+2$ is $w_0$. If $m$ is odd, a similar argument
applies.

For $j = 0,\dots,m+1$, let $u_j$ (respectively $u_j^\prime$) be the value of
$r_j$ in $V$ (respectively $V^\prime$) at time $m+j+1$. Then by induction on
$j$, $u_j^\prime = \neg u_j$. But then $w_0 = u_{m+1}^\prime = \neg u_{m+1}
= \neg w_0$, contradiction. Therefore $V^\prime$ must enter a state cycle
of size $\ge d$ when started in state $x$.

To summarize, we have shown that with probability $1 - o(1)$, there is some
gate in $R$, say $r_0$, such that $f_{r_0} \in {\cal C}$, and $V$ is strong
when $r_0$ is assigned one of the functions in $\cal C$. By symmetry,
the two choices are equally likely, and the Lemma follows.
\end{pf}
\subsection{Combinatorial Lemmas}
We now derive lower bounds on the probability of existence of sets of
vortices of various circumference. Let $D_n \subseteq [\beta\log n,
2\beta\log n]$ and $|D_n| = k(n)$ for each positive integer $n$. Our goal
is to find an asymptotic estimate for the probability that $\tilde{B}$
has strong vortices of circumference $d$, for all $d \in D_n$.
The approach is based on
sieve methods that are extensions of Ch. Jordan's formula
and Bonferroni's inequality. The monograph of Bollob\'{a}s \cite{b}
contains an exposition of these formulas. The extensions that we will
use are described in full generality in Lynch \cite{l.fun}.

Fixing $n$, put $k = k(n)$ and index the elements of $D_n$ by
$d_1,\dots,d_k$. For each $i = 1,\dots,k$ let ${\cal B}_i$ be an
indexed set of all subsets of $\{1,\dots,n\}$ of size $2d_i$, say
${\cal B}_i = \{C_{ij} : 1 \le j \le \binom{n}{2d_i}\}$.
For each $C_{ij}$ let ${\cal P}_{ij}$ be the property
``$B$ has a strong vortex of circumference $d_i$ on $C_{ij}$."

Take any family of sets
$$
\vec{S} = \{ S_i : 1 \le i \le k \}
$$
such that $S_i \subseteq {\cal B}_i$.
Let
$$
E^\ge(\vec{S}) = \bigcap_{i=1}^k
\biggl(\bigcap_{C_{ij} \in S_i} {\cal P}_{ij}\biggr).
$$
That is, $E^\ge(\vec{S})$ is the set of boolean cellular automata on
$n$ gates that have strong vortices on $C_{ij}$ for each $C_{ij} \in S_i$,
$i = 1,\dots,k$. Let $\vec{s} = \langle s_i : 1 \le i \le k \rangle$
be a sequence of positive integers and
$$
L(\vec{s}) = \sum\begin{Sb} \vec{S} \\ |S_i| = s_i \end{Sb}
  \pr(E^\ge(\vec{S}) | \tilde{B} \text{ is simple}).
$$
We put $\sum(\vec{s})$ for $\sum_{i=1}^k s_i$ and $\langle r \rangle^k$
for $\langle r,\dots,r \rangle$, the sequence of $k$ $r$'s, for any real
number $r$. We use $\vec{s} \ge \langle r \rangle^k$ to mean $s_i \ge r$
for $i = 1,\dots,k$. The next two lemmas are
applications of the extensions of Ch.
Jordan's formula and Bonferroni's inequality mentioned earlier.
\begin{lemma}\label{lemjordan}
We have
\begin{multline*}
\pr\left(\bigwedge_{i=1}^k \tilde{B}
\text{ has a strong vortex of circumference }
d_i | \tilde{B} \text{ is simple}\right) \\
= \sum_{\vec{s} \ge \langle 1 \rangle^k} (-1)^{\Sigma(\vec{s}) - k}
  L(\vec{s}).
\end{multline*}
\end{lemma}
\begin{lemma}\label{lembon}
For any $K \ge k$
$$
\sum \begin{Sb} \vec{s} \ge \langle 1 \rangle^k \\ \Sigma(\vec{s}) \ge K
\end{Sb} (-1)^{\Sigma(\vec{s}) - K} L(\vec{s}) \ge 0.
$$
\end{lemma}

The main result of this subsection is the next lemma.
\begin{lemma}\label{lemlbvortex}
Let $p_m$ be as given in Lemma
\ref{lemlocrec}, $k(n) = O(\log n /\log\log n)$, and
$\sigma_i$ be the probability that a vortex of circumference $d_i$
is strong, for $i = 1,\dots,k(n)$. Then
\begin{multline*}
\pr\left(\bigwedge_{i=1}^k \tilde{B}
\text{ has a strong vortex of circumference }
  d_i \wedge \tilde{B} \text{ is simple}\right)  \\
  = (1 - n^{-\Omega(1)})
\prod_{i=1}^k \left(1 - e^{-p_m^{d_i}\sigma_i}\right)
+ n^{-\Omega(\log\log n)}.
\end{multline*}
\end{lemma}
\begin{pf}
We will show that
\begin{multline*}
\pr\left(\bigwedge_{i=1}^k \tilde{B}
\text{ has a strong vortex of circumference }
  d_i | \tilde{B} \text{ is simple}\right) \\
  = (1 - n^{-\Omega(1)})
\prod_{i=1}^k \left(1 - e^{-p_m^{d_i}\sigma_i}\right)
+ n^{-\Omega(\log\log n)}.
\end{multline*}
The Lemma will follow by Lemma \ref{lemlbsimple}.
For $i = 1,\dots,k$ let $T_i$ be the set of all vortex types of
circumference $d_i$. Take any $\vec{S} = \{ S_i : 1 \le i \le k \}$
such that each $S_i \subseteq {\cal B}_i$, $|S_i| = s_i$, and
$C_{gh} \cap C_{ij} = \emptyset$ for all $(g,h) \neq (i,j)$, $C_{gh} \in S_g$,
$C_{ij} \in S_i$. By Lemma \ref{lemlocrec},
$$
\pr(E^\ge(\vec{S}) | \tilde{B} \text{ is simple}) =
  \prod_{i=1}^k \left[\sum_{\tau \in T_i} \frac{(2d_i)!}{\lambda(\tau)}
  \times \left(\frac1{\binom{n}2}\right)^{d_i}
  \times \left(\frac{p_m}2\right)^{d_i}
  \times \pi(\tau) \times \sigma_i \right]^{s_i}.
$$
By Lemmas \ref{lemlbrigid} and \ref{lemlbprob}, this is
$$
\prod_{i=1}^k \left[ (1 - n^{-\Omega(1)}) \left(\frac{p_m}{n(n-1)}\right)^{
  d_i}  \times (2d_i)! \times \sigma_i \right]^{s_i}.
$$
Then, using the falling factorial power notation $n^{\underline{k}} =
\prod_{i=0}^{k-1} (n-i)$,
$$
L(\vec{s}) = \frac{n^{\underline{\Sigma 2d_is_i}}}{\prod_{i=1}^k
  ((2d_i)!)^{s_i} s_i!}\times
  \prod_{i=1}^k \left[ (1 - n^{-\Omega(1)}) \left(\frac{p_m}{n(n-1)}\right)^{
  d_i}  \times (2d_i)! \times \sigma_i \right]^{s_i}.
$$

Let us approximate $L(\vec{s})$ when $\sum(\vec{s}) \le (\log n)^2$.
Since $1-x = e^{-x - O(x^2)}$ for $x \rightarrow 0$,
$\bigl(1 - n^{-\Omega(1)}\bigr)^{
\Sigma(\vec{s})} = 1 - n^{-\Omega(1)}$. Then, using Stirling's formula,
$$
L(\vec{s}) = \bigl(1 - n^{-\Omega(1)}\bigr) \prod_{i=1}^k
\frac{(p_m^{d_i}\sigma_i)^{s_i}}{s_i!}.
$$
The number of sequences $\vec{s}$ such that $\vec{s} \ge \langle 1
\rangle^k$ and $\sum(\vec{s}) = (\log n)^2$
is bounded above by
$$
\binom{(\log n)^2}{k-1} = \log n^{O(\log n/\log\log n)} = n^{O(1)}.
$$
For any such $\vec{s}$, there is some $i$ such that $s_i \ge \log n$.
Therefore
\begin{align*}
\sum\begin{Sb} \vec{s} \ge \langle 1 \rangle^k \\
  \Sigma(\vec{s}) = (\log n)^2 \end{Sb} L(\vec{s}) & =
  \frac{n^{O(1)}}{(\log n)!} \\
  & = n^{-\Omega(\log\log n)}.
\end{align*}
By Lemmas \ref{lemjordan} and \ref{lembon} (taking $K = (\log n)^2$),
\begin{align*}
\pr\left(\bigwedge_{i=1}^k \tilde{B} \right. &
\text{ has a strong vortex of circumference }
d_i | \tilde{B} \text{ is simple}\biggr) \\
  & = \bigl(1 - n^{-\Omega(1)}\bigr)\left[
  \sum\begin{Sb}\vec{s} \ge \langle 1 \rangle^k \\
  \Sigma(\vec{s}) \le (\log n)^2\end{Sb}
  (-1)^{\Sigma(\vec{s}) - k}
  \prod_{i=1}^k \frac{(p_m^{d_i}\sigma_i)^{s_i}}{s_i!} \right]
  + n^{-\Omega(\log\log n)} \\
  & = \bigl(1 - n^{-\Omega(1)}\bigr)\biggl[
  \sum_{\langle 1 \rangle^k \le \vec{s}
  \le \langle (\log n)^2 \rangle^k}
  (-1)^{\Sigma(\vec{s}) - k}
  \prod_{i=1}^k \frac{(p_m^{d_i}\sigma_i)^{s_i}}{s_i!} \biggr]
  + n^{-\Omega(\log\log n)} \\
  & = \bigl(1 - n^{-\Omega(1)}\bigr)
  \prod_{i=1}^k\biggl( \sum_{1 \le s
  \le  (\log n)^2}
  (-1)^{s - 1}
  \frac{(p_m^{d_i}\sigma_i)^{s}}{s!} \biggr)
  + n^{-\Omega(\log\log n)} \\
  & = \bigl(1 - n^{-\Omega(1)}\bigr)
  \prod_{i=1}^k\left(1
  -e^{-p_m^{d_i}\sigma_i} \right) + n^{-\Omega(\log\log n)}. \qed
\end{align*}
\renewcommand{\qed}{}\end{pf}
\begin{corollary}\label{corlbvortex}
If $k(n) = O(\log n/\log\log n)$, then
$$
\pr\left(\bigwedge_{i=1}^k \tilde{B} \text{ has a vortex of circumference }
  d_i \wedge \tilde{B} \text{ is simple}\right) =  n^{-o(1)}.
$$
\end{corollary}
\begin{pf}
By Lemma \ref{lemlocstab}
$$
p_m^{d_i} \ge \left(1 - \frac1{a\alpha\log n}\right)^{2\beta\log n}
\sim e^{-2\beta/(a\alpha)},
$$
and by Lemma \ref{lemlbstrong}, $\sigma_i \ge 1/2 - o(1)$.
Therefore
\begin{align*}
\prod_{i=1}^k \left(1 - e^{-p_m^{d_i}\sigma_i}\right) & \ge \prod_{i=1}^k
  \frac{p_m^{d_i}\sigma_i}2 \\
  & \ge (e^{-2\beta/(a\alpha)}/5)^{O(\log n/\log\log n)}
  \text{ (any constant $> 4$ will do)} \\
  & = n^{-o(1)}. \qed
\end{align*}
\renewcommand{\qed}{}\end{pf}
\subsection{Completion of Proofs}
\paragraph{\em Proof of Theorem \ref{thmmain1}}
For each $n$ let $D_n$ be the set of primes in $[\beta\log n,2\beta\log n]$.
By the Prime Number Theorem \cite{t},
$$
k(n) \sim \frac{\beta\log n}{\ln\log n}.
$$
Therefore by Corollary \ref{corlbvortex},
$$
\pr\left(\bigwedge_{i=1}^k \tilde{B}
\text{ has a strong vortex of circumference }
  d_i \wedge \tilde{B} \text{ is simple}\right) =  n^{-o(1)}.
$$

Take any $\tilde{B}$ satisfying the above condition. Since $\tilde{B}$
is simple, with probability $\ge 2^{-k(n)} = n^{-o(1)}$, a random starting
state takes each strong vortex of circumference $d_i$, $i = 1,\dots,k(n)$, to
a state cycle of size $d_i$ or $2d_i$. That is, for such a starting state,
$\tilde{B}$ enters a state cycle of size greater than or equal to
\begin{align*}
(\beta\log n)^{k(n)} & = e^{(1 - o(1))\beta\log n} \\
  & = n^{\beta\log e - o(1)}.
\end{align*}
Thus, with probability $\ge n^{-o(1)}$, $\tilde{B}$
enters a state cycle larger than $n^{\beta\log e - o(1)}$.
By Markov's inequality,
$$
\ex(\rvc) \ge n^{\beta}.
$$
Since $\beta$ was arbitrarily large, the Theorem follows. \qed

\paragraph{\em Proof of Theorem \ref{thmmain2}}
Take $D_n$ as in the previous proof.
Fixing $n$, for $i = 1,\dots,k(n)$ let $\rvx_i$ be the indicator random
variable that is 1 if and only if $\tilde{B}$ has a strong vortex of
circumference
$d_i$, and $\rvx = \sum_{i=1}^{k(n)} \rvx_i$. Then, still assuming simplicity,
by Lemma \ref{lemlbvortex},
$$
\ex(\rvx_i)  = \bigl(1 - n^{-\Omega(1)}\bigr)\left(1 -
e^{-p_m^{d_i}\sigma_i}\right)
  + n^{-\Omega(\log\log n)}.
$$
Since $k(n) = \Theta(\log n/\log\log n)$ and
$1 - e^{-p_m^{d_i}\sigma_i} \ge
e^{-2\beta/(a\alpha)}/5$,
\begin{align*}
\ex({\rvx}) & \sim \sum_{i=1}^{k(n)} 1 - e^{-p_m^{d_i}\sigma_i} \\
  & \rightarrow \infty.
\end{align*}
Similarly,
\begin{align*}
\ex(\rvx^2) & = \sum_{i=1}^{k(n)} \ex(\rvx_i) + 2\sum_{1 \le i < j \le k(n)}
  \ex(\rvx_i \rvx_j) \\
  & \sim \sum_{i=1}^{k(n)} 1 - e^{-p_m^{d_i}\sigma_i}
  + 2\sum_{1 \le i < j \le k(n)} \left(1 - e^{-p_m^{d_i}\sigma_i}\right)
  \left(1 - e^{-p_m^{d_j}\sigma_j}\right) \\
  & \sim (\ex(\rvx))^2.
\end{align*}
Therefore by Chebyshev's inequality, for any $\delta < 1$,
\begin{align*}
\pr(\rvx \le \delta\ex(\rvx) | \tilde{B} \text{ is simple}) & \le
  \frac{\ex(\rvx^2) - (\ex(\rvx))^2}{(1-\delta)^2(\ex(\rvx))^2} \\
  & \rightarrow 0.
\end{align*}
That is, almost all $\tilde{B}$ have at least
$\delta k(n)e^{-2\beta/(a\alpha)}/5$
strong vortices of distinct prime circumferences in
$[\beta\log n,2\beta\log n]$.
For all such automata, with probability $\ge
2^{-\delta k(n)e^{-2\beta/(a\alpha)}/5} = n^{-o(1)}$,
the starting state leads to a
state cycle larger than or equal to
\begin{align*}
(\beta\log n)^{\delta k(n)e^{-2\beta/(a\alpha)}/5} & \ge
  e^{\beta\delta e^{-2\beta/(a\alpha)}\log n/5} \\
  & = n^{\beta\delta e^{-2\beta/(a\alpha)}\log e/5}.
\end{align*}
By Markov's inequality,
$$
\ex(\rvc|\langle \tilde{D},\tilde{F} \rangle) \ge
n^{\beta\delta e^{-2\beta/(a\alpha)}\log e/5 - o(1)},
$$
and we can take any $\gamma < \beta\delta e^{-2\beta/(a\alpha)}\log e/5$.
In fact, as noted in Corollary \ref{corlbvortex}, the 5 can be
replaced by 4. \qed

Note that $\beta e^{-2\beta/(a\alpha)}$ has a unique maximum
when $\beta = a\alpha/2$. Therefore, since the only restrictions on
$\alpha$, $\beta$, and $\delta$ are that $\alpha < 1/2$,
$\alpha < \beta$, and $\delta < 1$, the
$\gamma$ in Theorem \ref{thmmain2} can be arbitrarily close to
$e^{-2/a}\log e/8$.
\section{Discussion}
As mentioned in the Introduction, there have been many computer
simulations of random boolean cellular automata, specifically the
uniform distribution model where $a = c = 1/8$. The results indicate
a rather slow, even sublinear, growth rate of the average state cycle
size as a function of the number of gates. At first glance, the
superpolynomial average size of state cycles given by Theorem
\ref{thmmain1} seems to contradict the experimental evidence.
There are two possible resolutions to this. First, $a = c$ is the
border were large state cycles are just beginning to appear. This may
not be noticeable until the number of gates is quite large. Perhaps
the simulated automata were not large enough.

Second, our proof shows that the large average state cycle size is due
to a small fraction of the automata that have very large state cycles.
It may be that most of the automata have relatively small state cycles.
Our other main result (Theorem \ref{thmmain2}) is consistent with
this. It gives a $n^\gamma$ lower bound on state cycle size averaged
over all inputs, for almost all networks $\langle D,F \rangle$.
The exponent $\gamma$ is quite small. For $a = c = 1/8$,
it is less than $2 \times 10^{-8}$. Two relevant open problems are
to improve the lower bound in Theorem \ref{thmmain2} and the upper
bound for state cycle size in \cite{lc}.

Other computer experiments indicate that systems on the edge of chaos
show complex computational capability. To formalize this notion in
terms of the model in this article, we should consider random boolean
cellular automata with inputs and outputs. Then, instead of looking at
stability measures, we should try to determine the conditions that
result in substructures that compute complex functions. If the
experimental evidence is correct, then the $a = c$ threshold
is the region where these substructures arise. The techniques used
here to prove the existence of large vortices may be applicable.

The model studied in this article is essentially a metaphore for complex
biological systems. Future work in this area will inevitably lead to
models with more biological detail and accuracy. Whether such models
will be mathematically tractable cannot be answered now, but there are
some simple generalizations of our model that may be pertinent to this
question. One example is random boolean cellular automata where the
probabilities of the functions assigned to gates do not necessarily
satisfy any symmetry conditions. An immediate question is whether the
results of \cite{l.rbca2} and this article extend to non-symmetric
probabilities. Another generalization is to random boolean cellular
automata whose gates need not have exactly two inputs. One-input
gates are just a special type of two-input gates, but the
population of three-input gates seems quite different because of the
large proportion of non-canalyzing functions.

Lastly, two technical problems are to analyze the stability of
random boolean cellular automata without constant gates, i.e.,
$a = 0$ and those where $a < c$. Results on the proportion of weak
gates indicate that $a < c$ is the chaotic region, but the proportion
of stable gates and nontrivial bounds on state cycle size are not
known. We make the following conjectures:
\begin{enumerate}
\item If $a < c$ then asymptotically $a/c$ of the gates are stable.
Recall that in this case, $a/c$ is the smaller of the fixed points
of the recurrence (\ref{eqrecp}).
\item As $a - c$ increases, stability of the system increases.
That is, the proportions of stable and weak gates increase, and
the size of the state cycle decreases.
\end{enumerate}

\end{document}